\documentclass[fleqn,usenatbib]{mnras}
\usepackage{amsmath}
\usepackage{txfonts}                
\usepackage{graphicx}               

\title[Channels M$_{\rm BH}$ growth on the mass - size plane]{Two channels of supermassive black hole growth as seen on the galaxies mass-size plane}

\author[D. Krajnovi\'c et al.]{
Davor Krajnovi\'c $^{1}$\thanks{E-mail:dkrajnovic@aip.de}, Michele Cappellari $^{2}$ \& Richard M. McDermid$^{3,4}$
\\
$^{1}$Leibniz-Institut f\"ur Astrophysik Potsdam (AIP), An der Sternwarte 16, D-14482 Potsdam, Germany\\
$^{2}$Sub-department of Astrophysics, Department of Physics, University of Oxford, Denys Wilkinson Building, Keble Road, Oxford OX1 3RH, UK\\
$^{3}$Department of Physics and Astronomy, Macquarie University, North Ryde, NSW 2109, Australia\\
$^{4}$Australian Astronomical Observatory, PO Box 915, Sydney NSW 1670, Australia\\
}

\date{Accepted XXX. Received YYY; in original form ZZZ}

\pubyear{2017}

\begin{document}
\label{firstpage}
\pagerange{\pageref{firstpage}--\pageref{lastpage}}
\maketitle

\begin{abstract}
We investigate the variation of black hole masses (M$_{\rm BH}$) as a function of their host galaxy stellar mass (M$_\ast$) and half-light radius ($R_e$). We confirm that the scatter in M$_{\rm BH}$ within this plane is essentially the same as that in the M$_{\rm BH} - \sigma$ relation, as expected from the negligible scatter reported in the virial mass estimator $\sigma_v^2=G\times$M$_\ast/(5\times R_e)$. All variation in M$_{\rm BH}$ happens along lines of constant $\sigma_v$ on the (M$_\ast$, $R_e$) plane, or M$_\ast \propto R_e$ for M$_\ast \lesssim2\times10^{11}$ M$_\odot$. This trend is qualitatively the same as those previously reported for galaxy properties related to stellar populations, like age, metallicity, alpha enhancement, mass-to-light ratio and gas content. We find evidence for a change in the M$_{\rm BH}$ variation above the critical mass of M$_{\rm crit}\approx2\times10^{11}$ M$_\odot$. This behaviour can be explained assuming that M$_{\rm BH}$ in galaxies less massive than M$_{\rm crit}$ can be predicted by the M$_{\rm BH} - \sigma$ relation, while M$_{\rm BH}$  in more massive galaxies follow a modified relation which is also dependent on M$_\ast$ once M$_\ast >$ M$_{\rm crit}$. This is consistent with the scenario where the majority of galaxies grow through star formation, while the most massive galaxies undergo a sequence of dissipation-less mergers. In both channels black holes and galaxies grow synchronously, giving rise to the black hole - host galaxy scaling relations, but there is no underlying single relation that is universal across the full range of galaxy masses. 

\end{abstract}




%
%

\section{Introduction}
\label{s:intro}

The relationship between supermassive black holes and their host galaxies continues to generate much interests in the literature. Since the discovery of the relationship between the mass of the black holes (M$_{\rm BH}$) and the galaxy luminosity \citep[$L$,][]{1995ARA&A..33..581K}, it was clear that these objects must be linked in an intricate way \citep[for a review of the development of ideas about the scaling relations between black holes and galaxies see][]{2016ASSL..418..263G}. The subsequent discoveries of the relations between M$_{\rm BH}$ and galaxy mass \citep[M$_\ast$,][]{1998AJ....115.2285M}, velocity dispersion \citep{2000ApJ...539L..13G, 2000ApJ...539L...9F}, circular velocity \citep{2002ApJ...578...90F}, and galaxy concentration \citep{2001ApJ...563L..11G, 2003RMxAC..17..196G}, as well as several secondary scaling relation \citep{2010ApJ...720..516B, 2011MNRAS.410.2347H, 2008ApJ...678L..93S}, only deepened the interest. These relations, their tightness and the dynamic range covering several orders of magnitude, indicate that the growth of supermassive black holes in the centres of galaxies, and galaxies themselves, must be closely related. The hope is that by understanding the properties of these relations, their universality, shape, tightness and related uncertainties, will also highlight and untangle the relevant processes that are involved in the growth of black holes and galaxies. 

The number of galaxies with measured black hole masses has dramatically increased over the last ten years \citep{2005SSRv..116..523F, 2013ARA&A..51..511K}, approaching a hundred estimates of M$_{\rm BH}$ based on dynamical models of stellar or gaseous motions \citep{2016ApJ...818...47S}. Once these are combined with estimates based on the reverberation mapping for active galactic nuclei (AGN) and upper limit measurements, the sample comprises more than 200 galaxies \citep{2016ApJ...831..134V}. Such numbers, while not yet of sufficient size for pure statistical studies, allow a more complex analysis of the scaling relations, specifically to investigate which of the various relation has the smallest scatter (and therefore is more fundamental), if the data actually support two (or more) power-law relations, and if there is a third parameter which could make the scaling relations even tighter. The main limitation of the sample, next to the relatively low number of galaxies, is that it does not represent the complete population of galaxies \citep{2007ApJ...660..267B}. The sample is biased towards bright (massive), nearby early-type galaxies (ETGs), with, possibly more massive black holes than in the average of the population, as these are easier to directly probe with dynamical models given the spatial resolution achieved by observations \citep{2010ApJ...711L.108B,2016MNRAS.460.3119S}.

Indications of non-universality of M$_{\rm BH}$ scaling relation come from the demographics of host galaxies, for example by investigating if spiral galaxies, galaxies with bars, pseudo-bulges or AGNs satisfy the same relations as more massive early-type galaxies \citep[e.g.][]{2006MNRAS.365.1082W,2009ApJ...695.1577G,2013ApJ...764..184M}. Although samples of non-ETGs galaxies are relatively small \citep[e.g.][]{2016ApJ...818...47S}, there are clear indications that they do not necessary follow the same scaling relations as the more massive ETGs. Barred galaxies, galaxies hosting masers or AGNs, or pseudo-bulges seem to be offset from the main relation \citep{2008ApJ...680..143G, 2008MNRAS.386.2242H, 2009ApJ...698..812G, 2010ApJ...721...26G, 2011Natur.469..374K, 2014MNRAS.441.1243H,2016ApJ...826L..32G}. Scaling relations for early-type and spiral galaxies are also offset with respect to each other \citep{2013ApJ...764..184M}. It is, however, difficult to establish black hole scaling relations based on morphological classifications, as the classification can be difficult \citep[e.g. recognising pseudo-bulges,][]{2015HiA....16..360G}, and the samples are small, or span a limited range in both M$_{\rm BH}$  and galaxy stellar mass.

Nevertheless, there are several arguments that offer a tantalising indications that the M$_{\rm BH}$ scaling relation are not universal. \citet{2007ApJ...662..808L} showed that the predictions from M$_{\rm BH} - \sigma$ relation and M$_{\rm BH} - L$ are different for high mass galaxies and brightest cluster galaxies in particular. The main issue is that the M$_{\rm BH} - \sigma$ relation, using measured velocity dispersions of galaxies, predicts M$_{\rm BH}$ that are rarely larger than a few times $10^9$ M$_\odot$, while the relation with the luminosity predicts M$_{\rm BH}$ in the excess of $10^{10}$ M$_\odot$. The origin of this tension is in the curvature of the $L - \sigma$ relation, which for the most luminous systems departs from the \citet{1976ApJ...204..668F} relation; galaxies with velocity dispersion larger than about 300 km/s are very rare \citep{2003ApJ...594..225S,2006AJ....131.2018B}. Crucially, as \citet{2007ApJ...662..808L} pointed out, there is a difference in $L - \sigma$ relations for galaxies with and without cores in their nuclear profiles. The relation is much steeper for core galaxies, following $\sim\sigma^7$, compared to canonical  $\sim\sigma^4$. This is true regardless of using the ``Nuker'' \citep{1995AJ....110.2622L} or (core-) Sersic \citep{2003AJ....125.2951G} parametrisation of the nuclear profiles \citep{2013ApJ...769L...5K}. 

More recently, the data from complete ATLAS$^{3D}$ survey \citep{2011MNRAS.413..813C} revealed that the previously reported major break in the $L-\sigma$ relation is not related to the transition between core and core-less galaxies. Instead, the break is clearly observed, consistently in both the $M_\ast-R_e$ and $M_\ast-\sigma$ relations, around a mass $M_\ast\approx3\times10^{10}$ M$_\odot$, and is present even when all core galaxies are removed \citep{2013MNRAS.432.1862C}. It appears related to the transition between a sequence of bulge growth and dry merger growth. However, a much subtler change \citep{2009MNRAS.394.1978H} in the $L-\sigma$ is observed around $M_\ast\approx2\times10^{11}$ M$\odot$. Crucially, this characteristic mass marks also a transition between (core-less) fast rotators and (core) slow rotators \citep{2013MNRAS.433.2812K}, indicating a transition in the dominant assembly process \citep[for a review see section 4.3 in][]{2016ARA&A..54..597C}.

Distinguishing between core and core-less galaxies in the black hole scaling relation is still of potentially great significance, as cores are predicted to be created by black holes. When massive galaxies (harbouring massive black holes) merge, their black holes will eventually spiral down to the bottom of the potential well and form a binary \citep{1980Natur.287..307B,1991Natur.354..212E}. The decay of the binary orbit will be enabled by removal of stars that cross it \citep{1996NewA....1...35Q, 2001ApJ...563...34M}, resulting in a central region devoid of stars, a core, compared to the initial steep power-law light profile. As the removed stars were mostly on the radial orbits, this process introduces a strong tangential anisotropy, significantly larger than that expected for an adiabatic black hole growth \citep{1995ApJ...440..554Q,2014ApJ...782...39T}. The mergers are dissipation-less (dry) and there is no significant star-formation which could refill the core. An implication of this effect is that there should also exist a relation between the M$_{\rm BH}$ and the size of the core region \citep{2007ApJ...662..808L}, or the missing stellar mass \citep{2004ApJ...613L..33G} in the most massive galaxies. The uncertainties in these relations are, however, not any smaller than in other relations, also because there is no unique way to measure the size of the core \citep[e.g.][]{2007ApJ...662..808L,2009ApJ...691L.142K, 2010MNRAS.407..447H, 2013AJ....146..160R, 2014MNRAS.444.2700D}.

Dividing galaxies into Sersic and core-Sersic (or power-law and core) is significantly different from looking at the difference between various morphological types (late or early-type galaxies, classical or pseudo-bulges, barred and non-barred, etc). A key point here is that the property on which the sample is divided is based on a physical process of the core scouring by black holes \citep{1997AJ....114.1771F}. Therefore, there is a working paradigm supporting a possibility of the non-universality of the M$_{\rm BH}$ scaling relations. This was explored by \citet{2012ApJ...746..113G}, \citet{2013ApJ...768...76S} and \citet{2013ApJ...764..151G}, who showed that galaxies with and without cores (using the Sersic and core-Sersic parametrisation) have different M$_{\rm BH}$ - M$_\ast$ relation. Adding AGNs (all with Sersic profiles), which extends the sample of galaxies to lower masses, gives even more support to such a break in the scaling relation \citep{2015ApJ...798...54G, 2015ApJ...813...82R}.  Another physically motivated separation of galaxies is to divide them into star forming and quiescent galaxies \citep{2016ApJ...830L..12T,2017arXiv170701097T}, prompted by the need to suppress star formation in galaxy evolution models, where the activity of central black hole provides a ready feedback mechanism. Furthermore, the expectation is that quiescent galaxies will have larger black hole masses. This is similar to diving galaxies into early- and late-types as done by \citet{2013ApJ...764..184M}, which also found evidence that early-type galaxies harbour more massive black holes. Neither of these studies, however, reported a break in the relations. These division are only approximately similar to Sersic/core-Sersic divisions as many early-type galaxies do not have cores, but are quiescent, and further work is needed to describe the shape of the scaling relations across galaxy properties.

Adding more parameters to the correlations with M$_{\rm BH}$ could, in principle, result in tighter relations. This was investigate by a number of studies over the past decade \citep[e.g.][]{2003ApJ...589L..21M,2007ApJ...665..120A, 2016ApJ...818...47S,2016ApJ...831..134V}. Such attempts, however, generally conclude that the decrease of scatter when adding an additional parameter is not substantial, and M$_{\rm BH} -\sigma$ is still considered the tightest and perhaps the most fundamental relation, inspite of the intrinsic scatter no-longer being considered consistent with zero \citep[e.g.][]{2009ApJ...698..198G}.

In this work we introduce a different approach. Instead of looking for the best scaling relation and then inferring the possible formation scenarios, we start from the emerging paradigm of the two phase formation of galaxies, from both a theoretical \citep{2010ApJ...725.2312O} and observational \citep{2013MNRAS.432.1862C,2015ApJ...813...23V} points of view \citep[as reviewed in][]{2016ARA&A..54..597C}. Assuming that black holes evolve in sync with galaxies, and are modified through similar processes, which are dependant on the galaxy mass and environment \citep[e.g.][]{2010ApJ...721..193P}, we search for the records of these processes in the dependance of black holes masses with galaxy properties. In particular, we consider the distribution of galaxies with black hole measurements in the mass - size diagram, an orthogonal projection of the thin Mass Plane \citep{2013MNRAS.432.1862C}. 

After defining the sample of galaxies with black hole masses that we will use (Section~\ref{s:obs}), we present the mass - size diagram and analyse its distribution of black hole masses (Section~\ref{s:mass-size}). In Section~\ref{ss:phot} we use the stellar photometry and sizes of latest compilation of galaxies black hole mass measurements from the literature \citep{2016ApJ...831..134V}, while in Section~\ref{ss:2mass} we repeat the exercise using 2MASS catalog values to estimate stellar masses and sizes, and provide a scaling relation simply based on 2MASS photometry alone. In Section ~\ref{s:discs}, we present a toy model that reproduces what is seen in the data and discuss the implication of our results, before concluding with a brief summary of main results (Section~\ref{s:con}).

%
%

\section{A compilation of black hole masses}
\label{s:obs}

\begin{figure}
\includegraphics[width=0.5\textwidth]{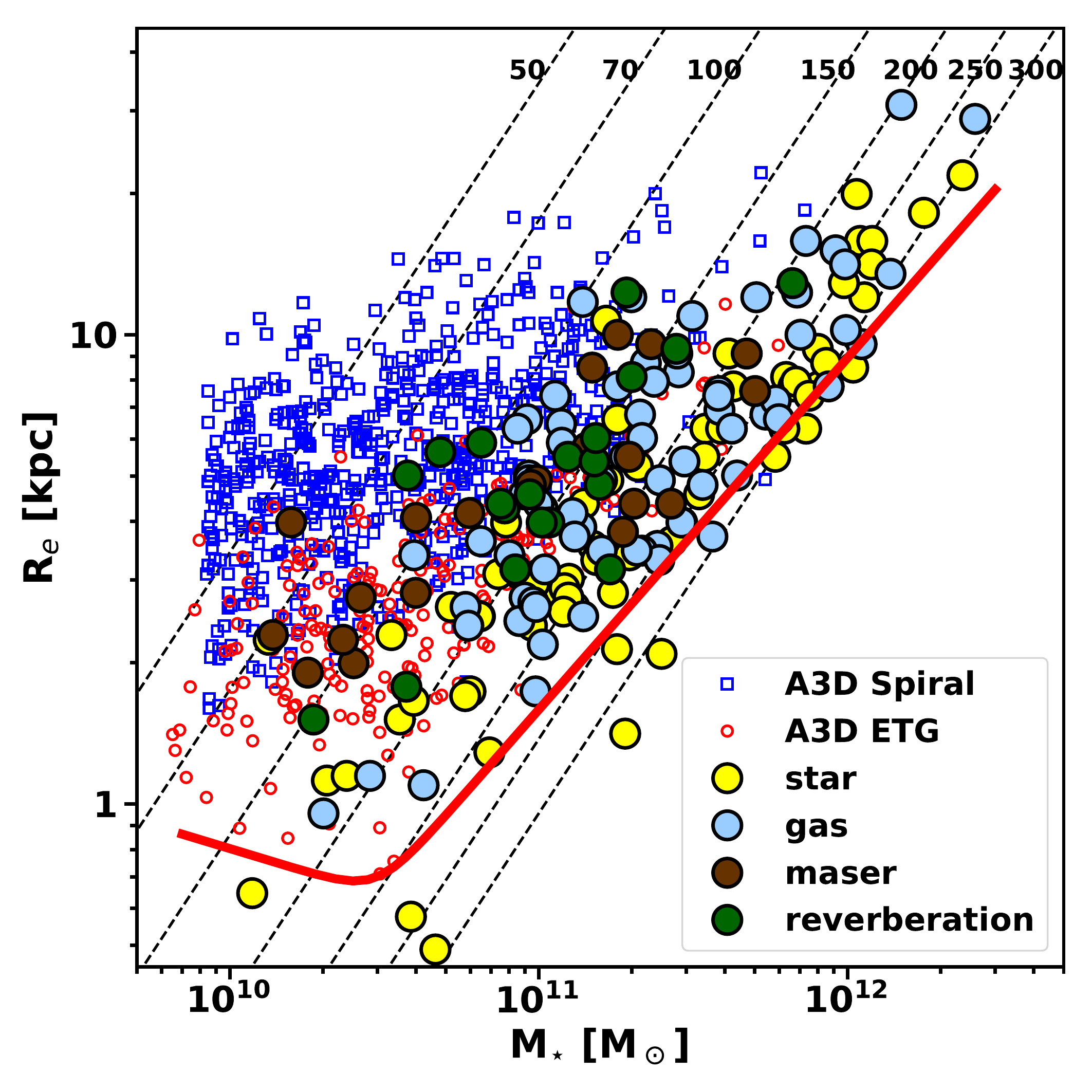}
\caption{Mass size diagram for galaxies with black hole mass measurements from the latest compilation presented in \citet[][see text for galaxies that were excluded]{2016ApJ...831..134V}. Filled circles are galaxies with black hole mass estimates, while their colour indicates the type of method used for black hole mass determination, as shown on the legend. Small open (red) circles and (blue) squares are ETGs and spirals from the ATLAS$^{\rm 3D}$ sample, respectively. Solid red line shows the "zone-of-exclusion" \citep[ZOE,][]{2013MNRAS.432.1862C} and the dashed-dotted diagonal lines show constant velocity dispersion, for values shown at the top of the diagram. Note that the mass and size estimates for the spirals, ETGs and the black holes sample have different origin, but the values for the black hole sample are internally consistent and uniformly measured. }
\label{f:ms1}
\end{figure}

We make use of the most recent compilation of M$_{\rm BH}$ measurements presented in Table 2 of \citet{2016ApJ...831..134V}, with black hole masses obtained from dynamical models and reverberation mapping. We do not include objects excluded from the regression fits in that paper, and we have also removed 49 upper limits. This results in 181 objects. Next to the black hole masses listed in \citet{2016ApJ...831..134V}, we also use the listed estimates for the size (effective or half-light radius, $R_e$) and the total $K_s$-band luminosity (Vega magnitudes) of these galaxies. We estimate the mass of these galaxies (M$_\ast$) using the relation between the $K_s$-band mass-to-light ratio and the velocity dispersion given by eq.~(24) in \citet{2016ARA&A..54..597C}. The velocity dispersions are also taken from \citet{2016ApJ...831..134V}. We stress that the velocity dispersion is the so-called {\it effective velocity dispersion} $\sigma_e$, which incorporates both the mean and random motions within an aperture the size of the effective radius. Unfortunately, the measurements of $\sigma_e$ are not uniform across the sample, as only for a subset of galaxies observed with integral-field units $\sigma_e$ can be measured directly. Figure~\ref{f:ms1} presents the mass - size diagram. Note that in the figure we do not plot three galaxies (NGC\,0221, NGC\,0404 and NGC\,4486B), which have significantly lower mass and sizes than the majority of the sample. We keep these objects for calculations, but do not show them for the presentation purposes. The only notable difference between this and the mass - size diagram in fig. 8 of \citet{2016ApJ...831..134V} is that the mass of the galaxies increased by about 15 per cent due to our slightly different conversion to mass. 

\begin{figure*}
\includegraphics[width=0.495\textwidth]{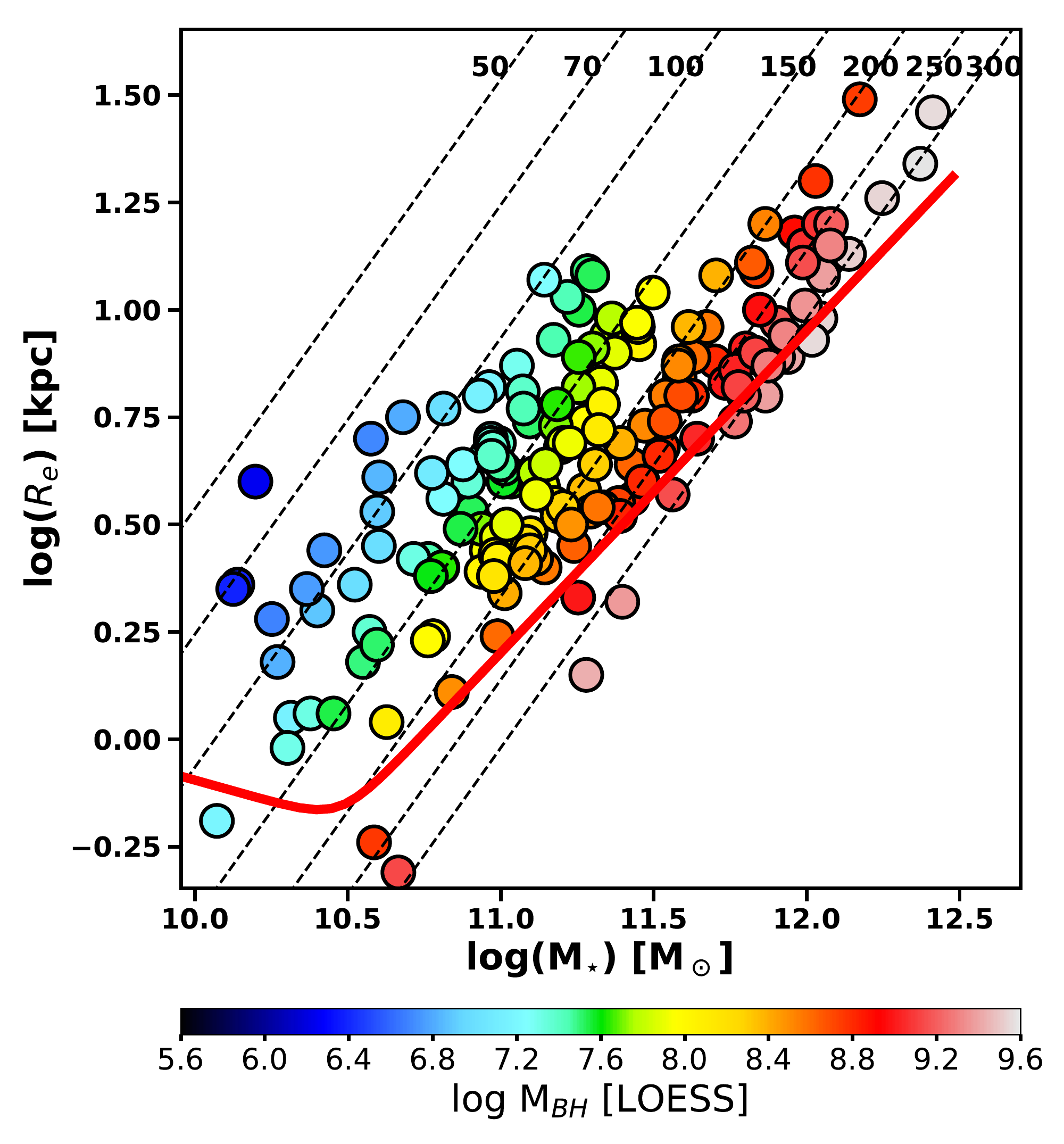}
\includegraphics[width=0.495\textwidth]{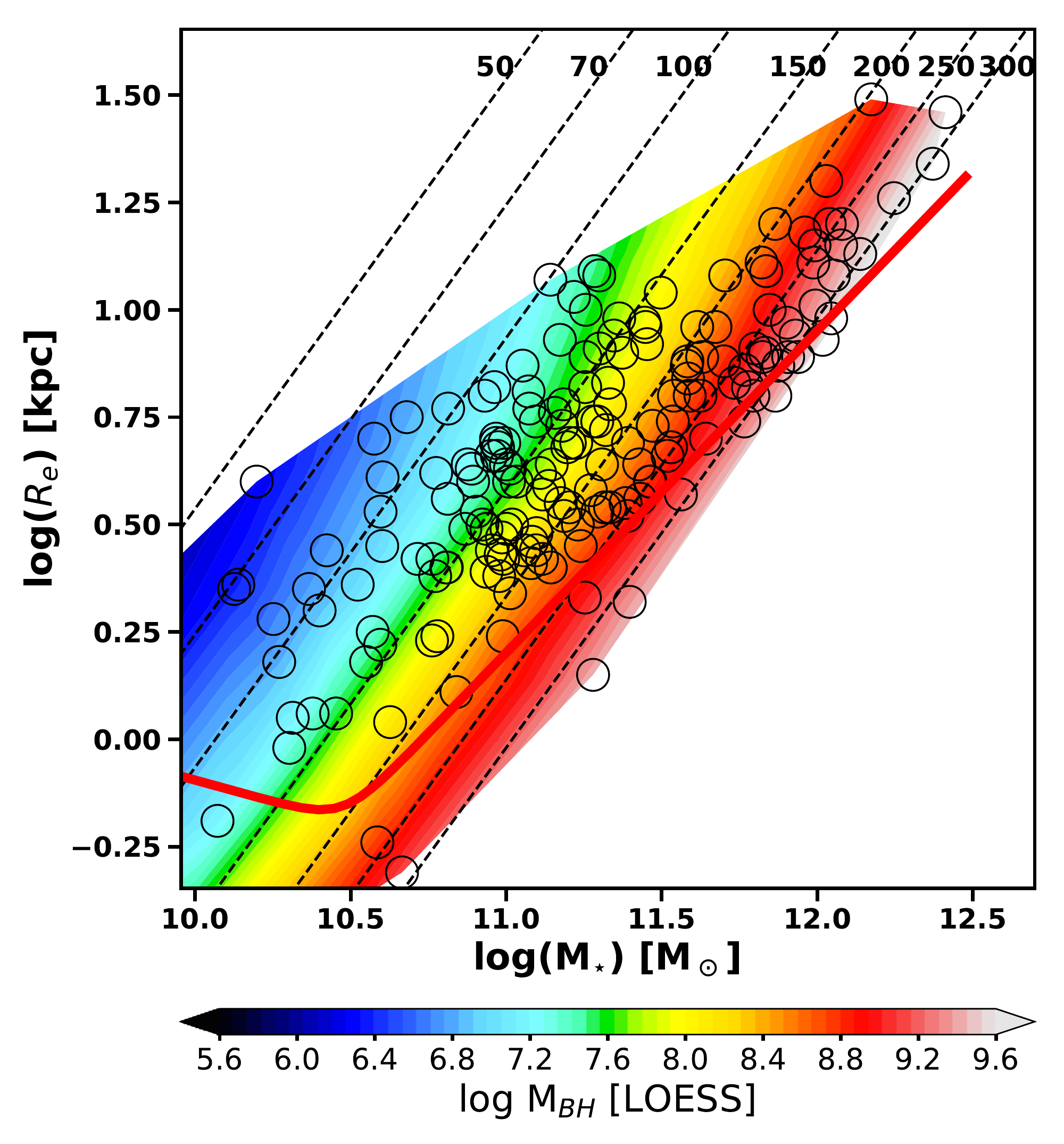}
\caption{Mass - size diagram as in Fig.~\ref{f:ms1}, but showing only galaxies with measured black hole masses. The colour of the symbols indicate the black hole masses within the range given on the colour-bar. M$_{\rm BH}$ values were smoothed using the LOESS method assuming constant errors. For an edge-on projection of the plot highlighting the scatter of M$_{\rm BH}$ see fig.~7 in \citet{2016ApJ...831..134V}. On the right panel, we show a continuous colour surface obtained by interpolating between the LOESS smoothed M$_{\rm BH}$ values. The red solid line is ZOE as on Fig.~\ref{f:ms1}. Diagonal dashed lines are lines of constant velocity dispersion. Note that the the contours depart from the constant velocity dispersion lines for masses M$_{crit}> 2\times 10^{11}$ M$_\odot$.}
\label{f:ms2}
\end{figure*}

In Fig.~\ref{f:ms1} we also plot galaxies from ATLAS$^{\rm 3D}$ survey, comprising early-type galaxies with masses measured via dynamical models \citep{2013MNRAS.432.1709C} and spiral galaxies from the parent sample. The mass estimates for the latter are based on their $K_s$ magnitudes following the eq.(2) from \citet{2013ApJ...778L...2C}. As already remarked by several authors \citep[e.g.][]{2016ApJ...831..134V}, galaxies with measured M$_{\rm BH}$ occupy a special place in this parameters space, typically being the most massive for a given radius and the smallest for a given mass. Inclusion of galaxies with M$_{\rm BH}$ measured using the reverberation mapping or galaxies with central masers, extends significantly the distribution towards low mass and low velocity dispersion regimes. The available M$_{\rm BH}$ determinations are also biased towards massive galaxies: many black hole masses have been measured in massive galaxies (e.g. $>5\times10^{11}$ M$_\odot$), where only a few such galaxies exist in the ATLAS$^{\rm 3D}$ volume limited sample. The black hole sample, while still relatively small and non-representative of the general galaxy population, spans a large range in the effective velocity dispersion (70 - 300 km/s), and is appropriate for the following analysis. We note that the results in this paper do not depend on the details of the photometric parameters, and we address this is Section~\ref{ss:2mass}.

%
%

\section{Black hole masses on the mass - size diagram}
\label{s:mass-size}

\subsection{Photometry from van den Bosch (2016)}
\label{ss:phot}

In the left panel of Fig.~\ref{f:ms2} we plot the mass - size diagram for galaxies with measured black hole masses. Now we also add M$_{\rm BH}$ as a third dimension shown by the colour. The data (M$_{\rm BH}$) are adaptively smoothed using the Locally Weighted Regression \citep[LOESS,][]{Clev:1979}. As shown in \citet{Clev:Devl:1988}, who also generalise the method to two dimensions, LOESS increases the visual information of a scatterplot  and can be used for data exploration, such as uncovering underlying trends which would be easier to observe in a much larger sample. In practice, we use the two-dimensional LOESS algorithm of \citet{Clev:Devl:1988}, as implemented in \citet{2013MNRAS.432.1862C}\footnote{Available from https://purl.org/cappellari/software}. We adopt a linear local approximation and a regularising factor f=0.5. To deal with different scale of the axes (log($R_e$) and log(M$_\ast$)), the software performs a robust estimation of the moment of inertia of the distribution and then performs a change of coordinates to transform the inertia ellipse into a circle. We assign a constant fractional error to all black holes and do not use the tabulated uncertainties on black hole masses, as they differ greatly from case to case, and generally ignore systematic uncertainties, which dominate the error budget. We can, however, confirm that the conclusions of this work do not change when using the reported uncertainties to weight the linear regression of the LOESS method.

The most striking feature of these plots is a change in black hole mass (LOESS smoothed and denoted by colour) from low to high (blue to read), which closely follows the diagonal lines of constant (virial) velocity dispersion. This is, of course, expected from the M$_{\rm BH} - \sigma$ relation, but the lines of constant velocity dispersion ($\sigma$) are actually predicted by the virial mass estimator $R_e=G\times$M$_\ast/(5\times \sigma^2)$, where G is the gravitational constant \citep{2006MNRAS.366.1126C}. To make the trend more obvious, we also interpolate across the region spanned by the galaxies from our sample, predicting the M$_{\rm BH}$ at every position in this plane. This results in a coloured surface plot as the background of the right panel of Fig.~\ref{f:ms2}. The interpolation was done based on the LOESS smoothed black holes masses of the sample galaxies.

In addition to highlighting the underlying trends, the LOESS method provides a nonparametric regression to a surface, which has the following significance: by using the LOESS smoothing we are, essentially, producing a non-parametric surface, which is defined locally. For the present purpose of investigating the relation between M$_{\rm BH}$ and other galaxy properties, this differs from what has been done previously. Other searches focused on defining a plane within the three-dimensional space of black hole mass, galaxy size and mass (or other parameters) \citep[e.g.][]{2003ApJ...589L..21M, 2016ApJ...818...47S,  2016ApJ...831..134V}. We are however, now not looking for one plane that best describes all parameters, but allow for a possible local bends of the plane, a change in the orientation of the plane within the space defined by $R_e$, M$_\ast$ and M$_{\rm BH}$. This LOESS fitted (non-parametric) surface is essentially shown by the contours in the right panel of Fig.~\ref{f:ms2}.

\citet{2013MNRAS.432.1862C} showed that the lines of constant velocity dispersion trace the mass concentration and the mass density (or bulge mass fraction) of galaxies below M$_{\rm crit} \approx 2\times10^{11}$ M$_\odot$. Therefore, Fig.~\ref{f:ms2} shows that M$_{\rm BH}$ behaves similarly to a variety of galaxy properties linked to the stellar populations, such as strength of H$\beta$ and Mg$b$ absorption, optical colour, molecular gas fraction, dynamical M/L, initial mass function normalisation, age, metallicity and $\alpha$-element abundance. This is summarised in \citet{2016ARA&A..54..597C}, using results from \citet{2011MNRAS.414..940Y}, \citet{2011ApJS..193...29A}, \citet{2013MNRAS.432.1709C}, \citet{2013MNRAS.432.1862C} and \citet{2015MNRAS.448.3484M}. We urge the reader to compare our Fig.~\ref{f:ms2} with fig. 22 from \citet{2016ARA&A..54..597C}. It is striking that for the majority of galaxies, their black hole masses follow the same trend as galaxy properties arising from star formation. This implies that the black hole growth is strongly related to the growth of the galaxy's stellar populations.

The adaptively smoothed mass-size diagram reveals another striking characteristic. At some point above $\approx2\times10^{11}$ M$_\odot$, the black hole masses cease to follow closely the lines of constant velocity dispersion. The iso-colour lines (the lines of constant M$_{\rm BH}$) change in slope from one that is the same to that of the iso-$\sigma$ lines, to a steeper one, more closely following the increase in mass. The change is gradual and subtle. It can be seen by following the change in the colour along the lines of constant velocity dispersion, for example for $\sigma = 200$ or 250 km/s. Along those lines, the symbol colours change from yellow to orange (for $\sigma=200$ km/s) and from orange to red (for $\sigma=250$ km/s) with increasing mass. This effect is more visible comparing the lines of constant velocity dispersion with the coloured contours\footnote{Note that for $\sigma<70$ km/s there is also a change in the shape of the contours, but this effect is based on 3-4 galaxies at the edge of the distribution and is not robust.} on the right panel of Fig.~\ref{f:ms2}. 

The effect indicates that beyond a certain galaxy mass, the black hole masses do not only follow the changes in velocity dispersion, but also changes in the galaxy mass. The detection of this transition is remarkable, especially when one considers that galaxies for which M$_{\rm BH}$ does not seem to follow iso-$\sigma$ lines closely, span only a factor of about 3-4 in galaxy mass. At a given velocity dispersion in that mass range, the observed range of black hole masses is approximately an order of magnitude, but taking into account a realistic factor of two in the uncertainties for black hole masses \citep[i.e. depending on the type of data and type of models used,][]{2006MNRAS.370..559S, 2009MNRAS.399.1839K, 2010MNRAS.401.1770V,2011ApJ...729..119G,2013ApJ...770...86W}, it is not surprising that the effect is marginal and difficult to see in the current data. Furthermore, the visualisation of the effect is hindered by the increasing closeness of iso-$\sigma$ lines and the scarcity of galaxies with masses M$_\ast>10^{12}$ M$_\odot$ and size $R_e>20$ kpc, as will be discussed later. Nevertheless, we will attempt to reproduce the effect by a simple model in the next section, but we first look for it using differently established luminosities and sizes of galaxies.

\subsection{Photometry from 2MASS All Sky Extended Source Catalog}
\label{ss:2mass}

\begin{figure*}
\includegraphics[width=0.49\textwidth]{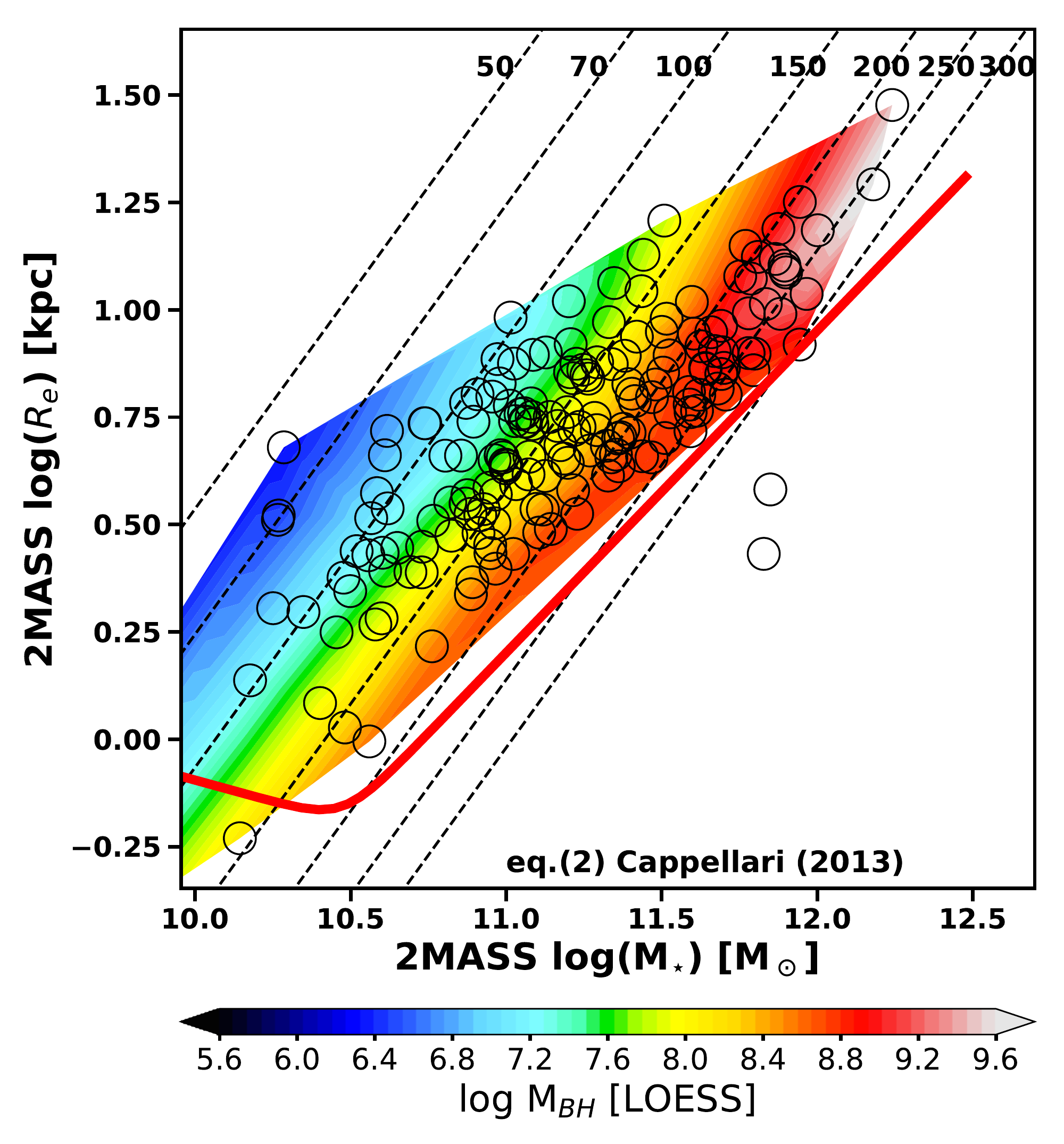}
\includegraphics[width=0.49\textwidth]{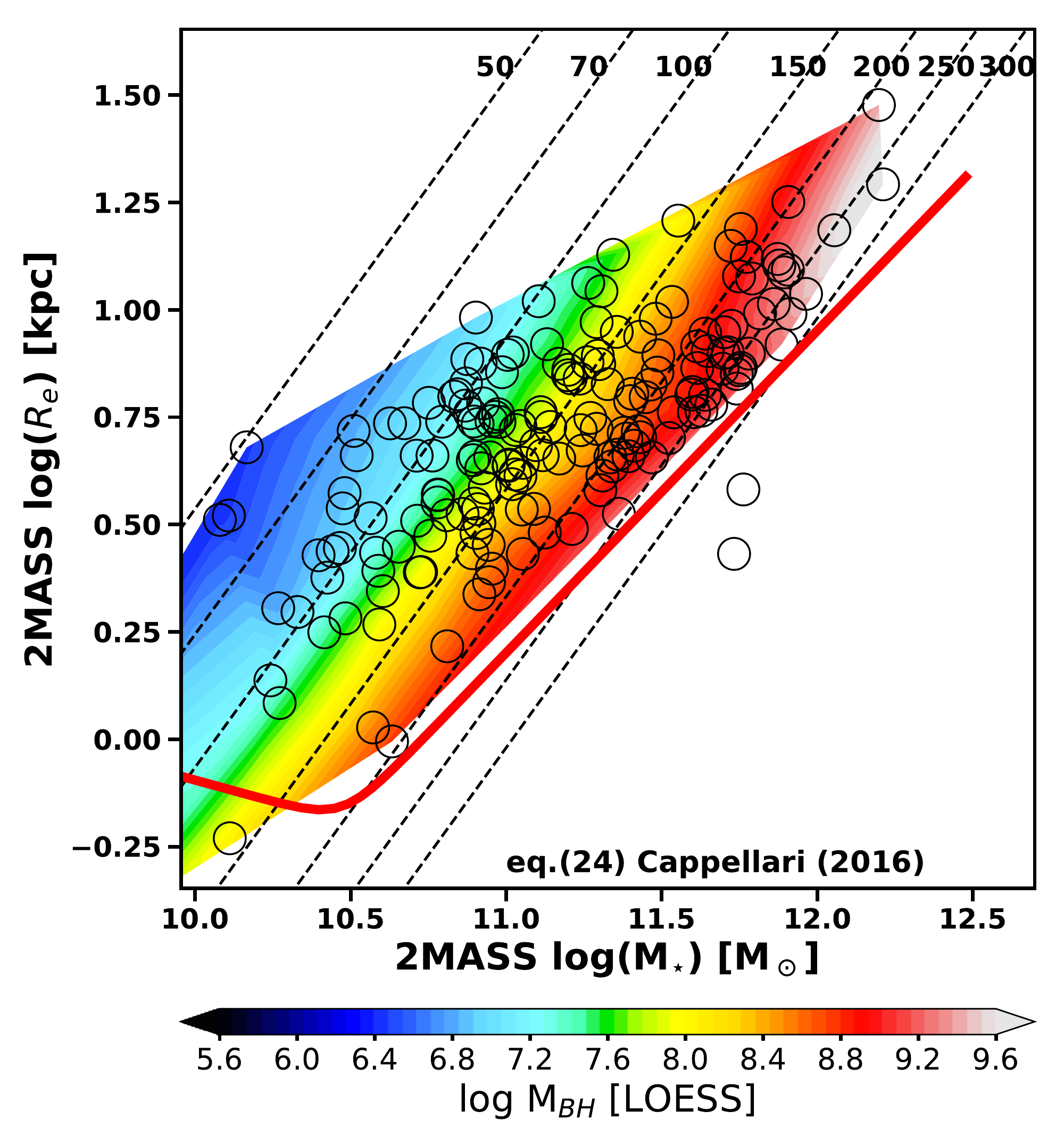}
\caption{Mass - size diagrams as in Fig.~\ref{f:ms2}, showing galaxies with measured black hole masses, with the galaxy luminosity and size obtained from the 2MASS All Sky Extended Source Catalog. The conversion to mass was achieved following two prescriptions based either on the mass - luminosity relation \citet{2013ApJ...778L...2C}, or the mass-to-light ratio - velocity dispersion relation \citet{2016ARA&A..54..597C}, shown on the left and right panels, respectively (see text for details). The symbols indicate the location of the galaxies in the mass - size plane. Their black hole masses were LOESS smoothed, and then interpolated into the continuous colour surface, showing the variation of the black hole masses in the mass - size plane, within the range given on the colour-bars. The red solid line is the ZOE. Two galaxies that fall strongly below ZOE, due to the contribution of the active nucleus to the total luminosity, were excluded from the LOESS fit. Diagonal dashed lines are lines of constant velocity dispersion. On both panels it is possible to see the change in the M$_{\rm BH}$ correlation from the velocity dispersion to mass, but the changes are not identical. This indicates the systematic error in the recovery of this effect from the choice of the photometry and conversion to mass.}
\label{f:2mass}
\end{figure*}

The location of galaxies on the mass - size plot depends on the global parameters, which for this study were obtained from \citet{2016ApJ...831..134V}. That study used 2MASS \citep{2000AJ....119.2498J} imaging and derived its own sizes and total K$_s-$band luminosities for all galaxies parameterising the surface brightness with \citet{1968adga.book.....S} profiles and using the growth curve approach. One of the reasons for this is that the 2MASS catalog values are typically found to underestimate the actual sizes and magnitudes. Furthermore, \citet{2016ApJ...831..134V} also used a detailed parametrisation of the point-spread function to account for the active galactic nuclei. Here we show that it is possible to qualitatively recover the same results by using 2MASS All Sky Extended Source Catalog \citep[XSC][]{2000AJ....119.2498J,2006AJ....131.1163S} data as they are. 

We query XCS for the size and extrapolated total magnitudes of galaxies from Table 2 of \citet{2016ApJ...831..134V}. Of 230 galaxies the search in XSC returned 228 sources from which we removed galaxies with upper limits on the black hole masses. We follow \citet{2013ApJ...778L...2C} and use the major axis of the isophote enclosing half of the total galaxy light in J-band (XSC keyword j\_r\_eff) which has better S/N than the K$_s-$band equivalent. Following \citet{2013MNRAS.432.1709C}, we define the size as $R_e = 1.61\times \rm j\_r\_eff$ and use the distances from \citet{2016ApJ...831..134V} to convert to physical units. 

Galaxy stellar masses (M$_\ast$) are estimated from the total magnitude (XSC keyword k\_m\_ext), in two different ways, based on the mass - luminosity and mass-to-light ratio  - velocity dispersion relations. In the {\it Approach 1}, we used the prescription from eq.~(2) of \citet{2013ApJ...778L...2C}, which relates the K$_s-$band absolute magnitude with the stellar mass, and is calibrated on the ATLAS$^{\rm 3D}$ sample of early-type galaxies, given their 2MASS K$_s-$band magnitudes and masses from \citep{2013MNRAS.432.1709C}. In the {\it Approach 2} we estimated the M/L ratio using eq.~{24} from \citet{2016ARA&A..54..597C} and the velocity dispersion from the compilation of \citet{2016ApJ...831..134V}. The obtained M/L is then multiplied with the $K$-band luminosity of galaxies $L_K$, which was obtained from the 2MASS magnitudes using the absolute magnitude of the Sun in K-band (M$_{\odot,K} = 3.29$) from \citet{2007AJ....133..734B}, as well as the distances from \citet{2016ApJ...831..134V}. These two approaches produce similar stellar masses, with the mean of the difference of the logarithms equal to 0.07 and the standard deviation of less then 0.1. 

In Fig.~\ref{f:2mass} we show the mass - size diagrams, equivalent to the right one in Fig~\ref{f:ms2}, but now using the sizes and the two different mass estimates described above. We perform the LOESS fit to both distributions, and compare them side by side. We exclude from the fit two galaxies which are significantly below the ZOE\footnote{These are Ark120 and Mrk0509, both known active galaxies with black hole estimates based on the reverberation mapping. Their 2MASS luminosities are likely biased.}. The circles show the distribution of the galaxies with measured black hole masses, while the underlying coloured surface represents the continuous variation of the black hole masses interpolated between the LOESS smoothed M$_{\rm BH}$. 

The panels in Fig.~\ref{f:2mass} differ from the right-hand panel of Fig.~\ref{f:ms2}, as galaxies have different mass and size measurements. However, panels in Fig.~\ref{f:2mass} show the same trends seen on Fig.~\ref{f:ms2}. M$_{\rm BH}$ values closely follow the lines of constant velocity dispersion for galaxies with masses below a few times $10^{11}$ M$_\odot$, following the behaviour or other properties of galaxies related to star formation \citep[as in fig~22 of][]{2016ARA&A..54..597C}. For highest mass galaxies the black hole masses deviate from the tight relation with the velocity dispersion\footnote{The deviation is also present for $\sigma<70$ km/s in the right panel, but it is due to the lack of data points for robust interpolation.}. The departure from a relation with velocity dispersion occurs above a mass a few times $10^{11}$ M$_\odot$. It is a remarkable fact that by using the ready catalog values one can produce the plot qualitatively similar to Fig.~\ref{f:ms2}. This adds weight to the robustness of the result presented in Section~\ref{ss:phot}. 

For completeness of this section and to provide ready values for readers interested in using XCS catalog, we present the best fit relation between M$_{\rm BH}$, stellar mass M$_\ast$ and the effective radius $R_e$. We use the {\it Approach 1} values shown on the left panel of Fig.~\ref{f:2mass}, based on the compilation of M$_{\rm BH}$ from \citet{2016ApJ...831..134V} and M$_\ast$ estimated using \citet{2013MNRAS.432.1709C} prescription, which relies only on the XSC values and distances from \citet{2016ApJ...831..134V}. We fit the relation of the form: 

\begin{equation}
\label{eq:2mass}
\begin{split}
\log \bigg( \frac {M_{\rm BH}}{10^8 M_\odot} \bigg)  = a + b \log \bigg(\frac{M_\ast}{10^{11} M_\odot} \bigg) + c \log \bigg(\frac{R_e}{5 kpc} \bigg),
\end{split}
\end{equation}

\begin{figure}
\includegraphics[width=0.5\textwidth]{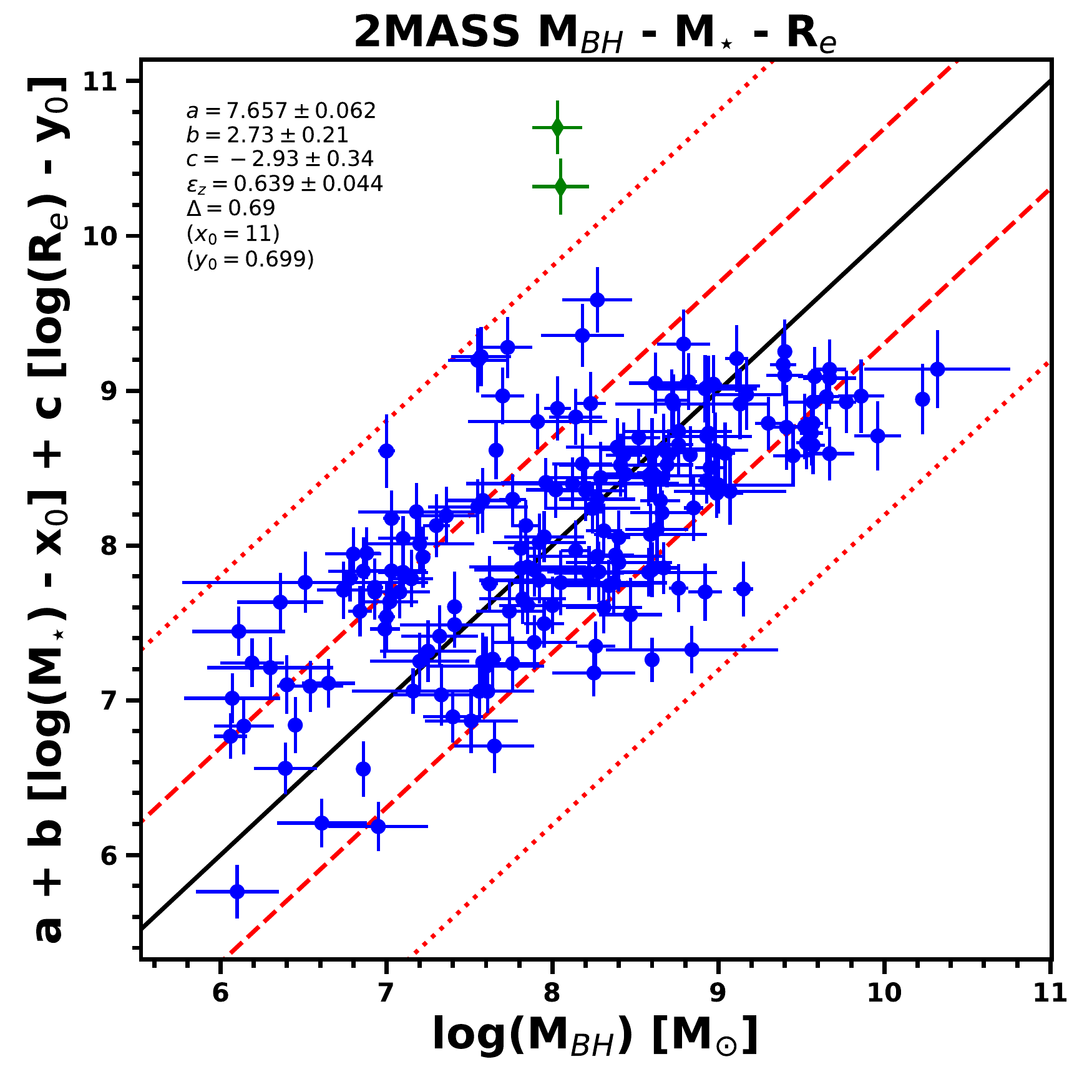}
\caption{Black hole mass - stellar mass - effective size relation with the best fit of the form as given by eq.~\ref{eq:2mass}. Photometric properties are taken from 2MASS XSC catalog and the mass is estimated using the mass - luminosity relation from \citet[][eq. (2)]{2013MNRAS.432.1709C} and distance from \citet{2016ApJ...831..134V}. Best fit parameters are given in the legend, where $\Delta$ is the root-mean-square ({\it rms}) scatter of the fit and the intrinsic scatter around the M$_{\rm BH}$ axis is given by $\epsilon_z$. Dashed and dotted lines are one and two times the {\it rms} scatter, respectively. Green symbols are data points rejected during the fit.  }
\label{f:fits}
\end{figure}

\noindent The fit was performed using the least trimmed squares fitting method LTS\_PLANEFIT\footnote{Available at http://purl.org/cappellari/software} of \citet{2013MNRAS.432.1709C}, which combines the Least Trimmed Squares robust technique of \citet{ROUSSEEUW2006} into a least-squares fitting algorithm which allows for errors in all variables and intrinsic scatter. We used the tabulated errors for M$_{\rm BH}$ from \citet{2016ApJ...831..134V}, while for galaxy sizes we follow \citet{2013MNRAS.432.1709C}, assuming 10 per cent errors, and we approximate the uncertainty on the mass to be of the order of 10 per cent. The fit is shown in Fig~\ref{f:fits}. As a consistency check, we also fitted the relation from the {\it Approach 2}, where the M$_\ast$ is obtained using eq.~24 of \citet{2016ARA&A..54..597C} and the velocity dispersion compilation from \citet{2016ApJ...831..134V}. The results can be seen in Table~\ref{t:fits}, and they are consistent between each other. Both fits are consistent with those of galaxy mass - size - black hole mass relation of \citet{2016ApJ...831..134V}. 

Scatters reported in Table~\ref{t:fits} are somewhat higher than those from recent detailed studies of the scaling relations \citep{2016ApJ...831..134V,2016ApJ...818...47S}. This is expected as we did not performed any cleaning of XSC values and use approximate errors. It is actually remarkable that the best fit parameters are consistent with estimates based on different photometry, and that even the scatters are comparable. Although the {\it Approach 2} fit has a smaller scatter, we, nevertheless, advise usage of the {\it Approach 1} relation, as it does not depend on still mostly uncertain velocity dispersion measurements for galaxies with black hole masses. 

\begin{table}
   \caption{Black hole scaling relations based on 2MASS XSC data.}
   \label{t:fits}
$$
  \begin{array}{c c c c c c }
    \hline
    \noalign{\smallskip}

$fit$		& a			&b			& c   &\epsilon_z & \Delta\\
   \noalign{\smallskip} \hline \hline \noalign{\smallskip}
$Approach $ 1     & 7.66 \pm 0.06 &  2.7 \pm 0.2 & -2.9 \pm 0.3 & 0.64 \pm 0.04 & 0.7\\
$Approach $ 2     & 7.77 \pm 0.05 & 2.6 \pm 0.2 & -2.7 \pm 0.3 & 0.54 \pm 0.04  & 0.6\\

       \noalign{\smallskip}
    \hline
  \end{array}
$$ 
{Notes -- The form of the relation is given in eq.~\ref{eq:2mass}, while $\epsilon$ is the intrinsic scatter around the M$_{\rm BH}$ axis of the plane and $\Delta$ is the root-mean square scatter. For {\it Approach 1}, the stellar mass was estimated using XSC k\_m\_ext keyword, distances from \citet{2016ApJ...831..134V} and the mass - luminosity relation from \citet[][eq. (2)]{2013MNRAS.432.1709C}. For {\it Approach 2}, the stellar mass was estimated using the mass-to-light ratio - velocity dispersion relation from  \citet[][eq. (24)]{2016ARA&A..54..597C}, the velocity dispersion compilation from \citet{2016ApJ...831..134V} and luminosity of galaxies, which was converted from XSC k\_m\_ext  values using the absolute magnitude of the Sun in K-band (M$_{\odot,K} = 3.29$) \citep{2007AJ....133..734B}, and distances from \citet{2016ApJ...831..134V}. For both fits, sizes were estimated using XSC keyword j\_r\_eff, as $R_e = 1.61\times \rm j\_r\_eff$.
}
\end{table}

%
%

\section{Discussion}
\label{s:discs}

\begin{figure*}
\includegraphics[width=0.49\textwidth]{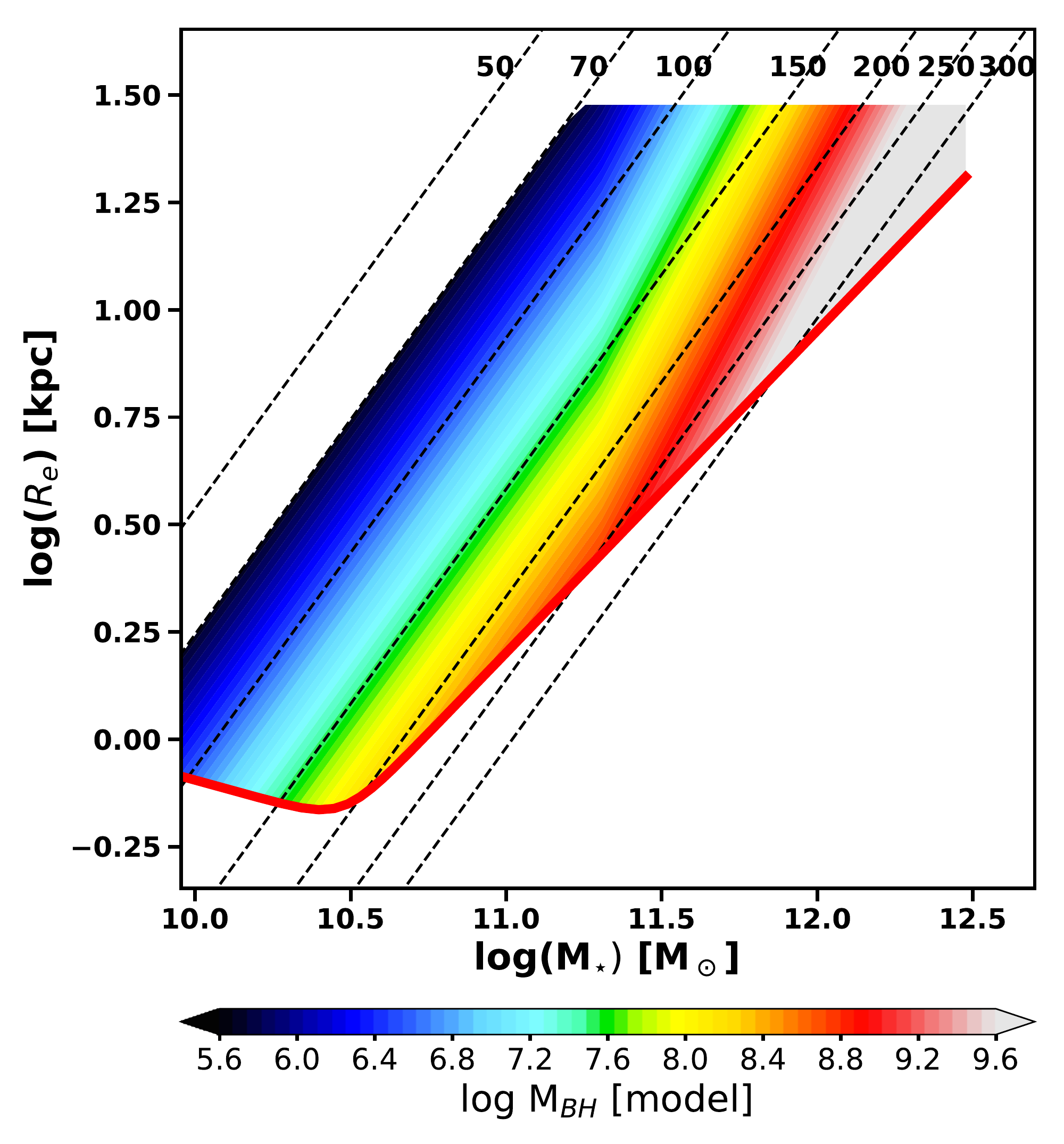}
\includegraphics[width=0.49\textwidth]{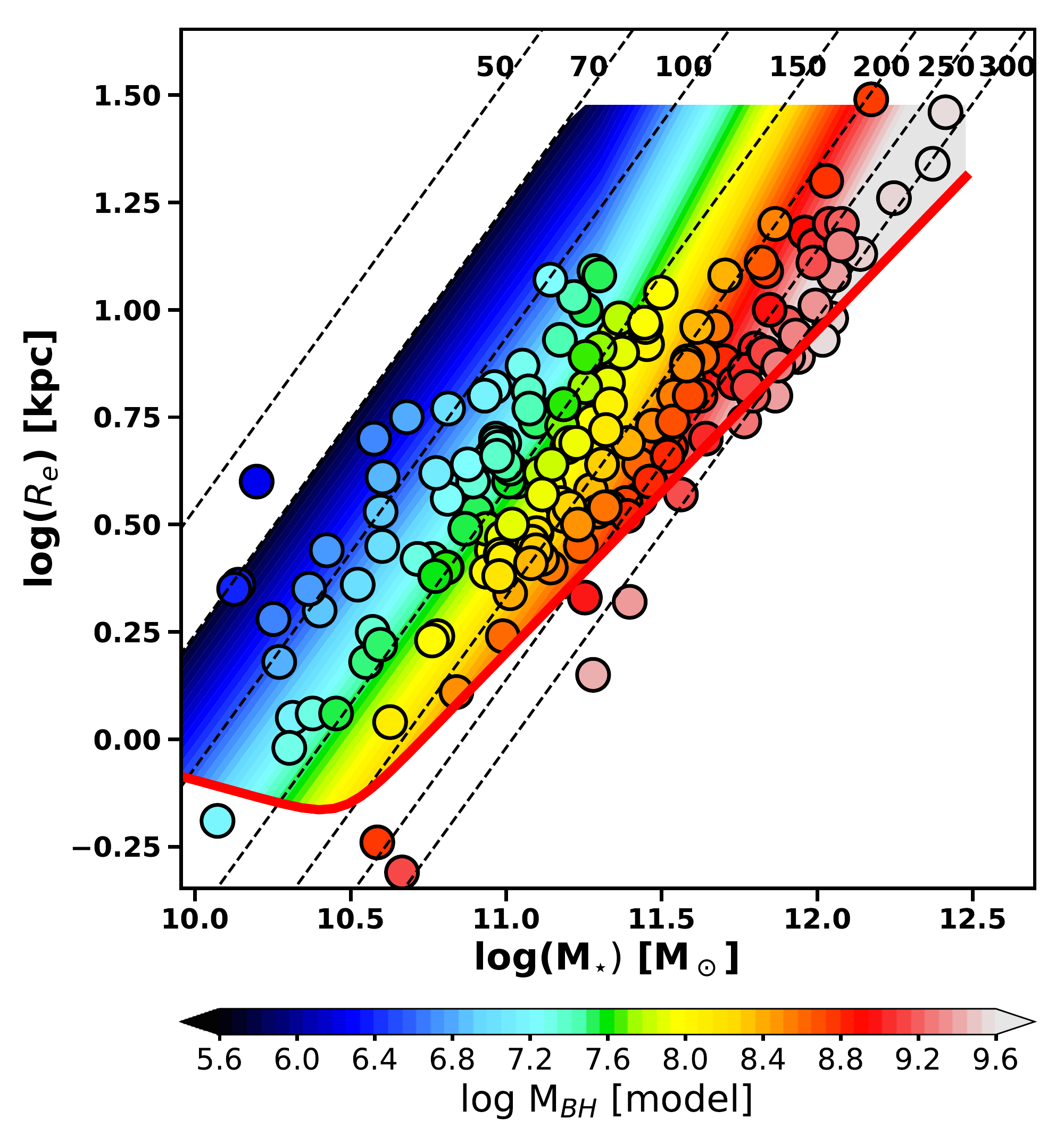}
\caption{A toy model simulation of M$_{\rm BH}$ on the mass - size plane (left) and the same model with over-plotted LOESS smoothed black hole masses (right). The toy model  M$_{\rm BH}$ are based on a simple prescription in which black hole masses are calculated based on the M$_{\rm BH} - \sigma$ relation for galaxies less massive than $2\times10^{11}$, and on a M$_{\rm BH} - \sigma$ modulated by the galaxy mass for higher mass galaxies (see eqs.~\ref{eq:sig} and~\ref{eq:mass}). The colour scale of the model is limited to the same range as the LOESS smoothed black hole masses of the sample galaxies, as shown on the colour bars. The red solid lines show ZOE. Diagonal dashed lines are lines of constant velocity dispersion. We also restricted the model to $\sigma>70$ km/s as there are no galaxies that could be compared with in that region.} 
\label{f:ms3}
\end{figure*}

\subsection{A toy model for evolution of black hole masses}
\label{ss:toy}

Here we want to understand what kind of signature we should expect to observe, based on simple assumptions on the growth process of supermassive black holes. We construct a toy model based on the assumption that below a critical mass of $2\times10^{11}$ M$_\odot$ the black hole mass can be predicted from the M$_{\rm BH} - \sigma$ relation, while above this mass the M$_{\rm BH} - \sigma$ relation is modulated by an additional term depending explicitly on the stellar mass. Our assumption follows the physically motivated distinction in black hole growth. Galaxies with masses less than $2\times10^{11}$ M$_\odot$ follow the channel of gas accretion, bulge growth and quenching, while more massive galaxies are descendents of galaxies formed in intense star-bursts at high redshifts and since than have grown following a channel of dissipation-less (dry) merging. 

We construct a toy model by defining a galaxy sample in a mass - size plane, spanning between 0.5 and 30 kpc in the effective radius and $10^9$ and $3\times10^{12}$ in mass, but restricted to be above ZOE. This distribution of points is similar to the one on Fig.~\ref{f:ms1}, including spiral galaxies, which occupy the large size - low mass part of the plane. The velocity dispersion of each galaxy is estimated using the virial relation $\sigma_e^2 = (M_\ast \times G)/(5\times R_e)$ \citep{2006MNRAS.366.1126C}. For each galaxy, depending on their stellar masses, we then calculate two sets of black holes masses using relations: 
\begin{equation}
\label{eq:sig}
\begin{split}
\log \bigg( \frac {M_{\rm BH}}{M_\odot} \bigg)  = \alpha + \beta \log \bigg(\frac{\sigma_e}{200 {\rm km/s}} \bigg), \\
{\rm for }~M_{\ast} < 2\times10^{11} M_\odot,
\end{split}
\end{equation}
and 
\begin{equation}
\label{eq:mass}
\begin{split}
\log \bigg( \frac {M_{\rm BH}}{M_\odot} \bigg)  = \alpha  + \beta \log \bigg(\frac{\sigma_e}{200 {\rm km/s}} \bigg) +\log \bigg(\frac{M_\ast}{2\times10^{11} M_\odot} \bigg),\\
{\rm for}~M_{\ast} > 2\times10^{11}  M_\odot
\end{split}
\end{equation}
\noindent where $\alpha=8.22$ and $\beta=5.22$ are taken from the M$_{\rm BH} - \sigma$ relation for all non-bared galaxies from \citet{2013ApJ...764..151G}. The exact choice of the M$_{\rm BH} - \sigma$ relation is not important. As we only assume that there is a trend with the velocity dispersion for low mass galaxies, we use a relation fitted to galaxies with typically low velocity dispersions and masses. In our toy model, M$_{\rm BH}$ varies smoothly from being a $\sigma$ dominated to being a M$_\ast$ dominated. Galaxies with mass less than $2\times10^{11}$ M$_\odot$ are fully in the regime of the M$_{\rm BH} - \sigma$ dependance. As the stellar velocity dispersion saturates above 300 km/s, the stellar mass eventually becomes the dominant contributor, but this happens significantly only for galaxies of the highest masses. 

We plot the results in Fig~\ref{f:ms3}, showing our model predictions in the mass - size plane. The colour of the contours specifies the variation of black hole masses.  We limit the colour scale to the extent for the LOESS smoothed values of the sample galaxies, which are over-plotted as coloured circles in the right panel. The colours specified by the model follow the trend similar to that in Fig.~\ref{f:ms2}. For low mass galaxies, the black hole mass follows the lines of constant velocities, but black hole masses in galaxies with M$_\ast>2\times10^{11}$ M$_\odot$ and $\sigma>200$ km/s start departing from the constant velocity dispersion lines, being higher than expected from the pure M$_{\rm BH} -\sigma$ relation. The predictive power of our toy model with no free parameters is surprising. It is best visible when the model contours are compared with the LOESS smoothed black hole masses of the sample galaxies (right panel of the figure), which closely follow the changes in the underlying colour (M$_{\rm BH}$) of the toy model.

The toy model also justifies the choice of the critical mass M$_{\rm crit}= 2\times10^{11}$ M$_\odot$ for the transition between the models. The choice is primarily physically motivated as this mass separates core-less fast rotators from core slow rotators \citep{2013MNRAS.433.2812K,2013MNRAS.432.1862C} and, therefore, it is the mass above which dissipation-less mergers (and likely black hole mergers) start dominating the galaxy (black hole) evolution. The change is gradual, reflecting the dominating contribution of the velocity dispersion even for galaxies above M$_{\rm crit}= 2\times10^{11}$ M$_\odot$, but also the expectation that only the most massive galaxies will experience a sufficient number of major dry mergers to significantly modify their black hole masses via direct black hole mergers. 

Fig.~\ref{f:ms3} also shows why the dependance of M$_{\rm BH}$ with M$_\ast$ for high masses is difficult to detect. As referred before, the current sample of galaxies with measured black hole masses is tracing only the edges of the galaxy distribution. The large region populated by spiral galaxies and low mass fast rotators is poorly explored. As it is not constrained, we limit the prediction of the toy model to $\sigma> 70$ km/s. 

For the detection of the bend in the properties of M$_{\rm BH}$ in the mass - size plane, however, there are two significant regions. The first one is centred on the large ($R_e>10$ kpc) and relatively less massive galaxies, around the line of constant velocity dispersion of about 150 km/s. In the nearby universe, where black hole masses can be measured with dynamical methods, these galaxies are rare and are typically spirals, as shown by the ATLAS$^{\rm 3D}$ Survey (see Fig~\ref{f:ms1}). The rarity of these systems currently exclude a possibility for a direct comparison with the proposed model, but further determination of black hole masses using molecular gas kinematics \citep[e.g.][]{2013Natur.494..328D, 2016ApJ...822L..28B, 2017MNRAS.468.4663O, 2017MNRAS.468.4675D} offer a possible route of exploring this range. 

The second region is the branch of the most massive galaxies, with sizes of 20 kpc or more, masses in excess of $10^{12}$ M$_\odot$ and $\sigma>250$ km/s. These galaxies are also rare in the nearby universe, but they are present in the form of brightest cluster galaxies or cD galaxies. In order to improve on the current description of the M$_{\rm BH}$ dependance, more galaxies of the highest masses and largest sizes need to be probed \citep[e.g.][]{2011Natur.480..215M, 2012ApJ...756..179M, 2016Natur.532..340T}, preferably through dedicated surveys \citep[e.g.][]{2014ApJ...795..158M}.

As mentioned above, the change in the M$_{\rm BH}$ dependance implies that the plane defined by M$_{\rm BH}$, M$_\ast$ and R$_e$, has a change of curvature at high masses. Its significance can be tested by the scatter from a fit defined by eqs.~\ref{eq:sig} and \ref{eq:mass}, in comparison with that from a standard M$_{\rm BH} - \sigma$ regression. We used a least squares method to fit a linear regression to eq.~\ref{eq:sig} for all galaxies plotted in Fig.~\ref{f:ms2} regardless of their mass. We then also repeated the same fit using eqs.~\ref{eq:sig} and~\ref{eq:mass} taking into account the mass dependance as described in the equations. We did not take into account the observed uncertainties on any variable, as our intention was not to find the best fit relation, but just to compare if there was a decrease in the scatter of the residuals. We compared the standard deviation of the residuals of the fits to these two equations and found that there was no improvement going from the first ($\Delta=0.533$) to the second ($\Delta=0.532$) fit. Equivalent results are obtained if instead of fitting we use for $\alpha$ or $\beta$ values derived in \citet{2016ApJ...831..134V}. Therefore, the model described by eqs.~\ref{eq:sig} and~\ref{eq:mass}, using the current database of black hole masses is not necessarily warranted in terms of providing a better correlation. This is not surprising given that, especially for lower mass galaxies, the total galaxy mass is not always found to be the best predictor of the black hole mass \citep[e.g.][]{2011Natur.469..374K,2016ApJS..222...10S,2016ApJ...818...47S}, although decomposing galaxies and using masses of certain components is difficult and uncertain \citep[e.g.][]{2014ApJ...780...70L, 2016ApJS..222...10S}.  \citet{2016ApJ...818...47S} find that introducing a bivariate correlations with bulge mass and velocity dispersion reduces the overall scatter, even when core ellipticals are removed from the scatter. On the other hand, \citet{2016ApJS..222...10S}, by decomposing galaxies observed at 3.6 $\mu$m and using bulge masses, were able to fit different scaling relations for early- and late-type galaxies. Our model, by concentrating on the total mass, however, remains physically motivated, describes the behaviour of the data in the mass - size diagram, and, crucially, it is easily testable with better samples of black hole masses which will likely follow with time.

\subsection{Origin for the non-universality of the M$_{\rm BH}$ scaling relations}
\label{s:theor}

The transition of the dependance of black hole mass from velocity dispersion (following the trends of other star formation related properties) to galaxy mass supports current ideas on the growth of galaxies and black holes \citep[e.g.][]{2013ApJ...768...76S}. The main processes regulating the growth of galaxies can be separated between those related to the in-situ star formation and those related to the assembly of galaxy mass by accretion of elsewhere created stars. Black holes also grow via two types of processes: by accretion of gas or by merging \citep{2010A&ARv..18..279V}. Accretion of material onto a black hole (i.e. gas originating from gas clouds or from destruction of passing stars), converts gravitational energy into radiation and results in active galactic nuclei, or quasars, when the radiation is particularly strong. Moreover, this process influences both the growth of the black hole and the galaxy. 

Current models predict that M$_{\rm BH}$ growth by accretion is proportional to $\sigma^4$ \citep{1999MNRAS.308L..39F,2005ApJ...618..569M} or $\sigma^5$ \citep{1998A&A...331L...1S, 1998MNRAS.300..817H}, establishing the relations with the host galaxies properties. The scatter of the M$_{\rm BH} - \sigma$ relation is too high to distinguish between these cases, partially due to difficulties in reducing the systematic errors in measuring black hole masses. However, at least part of the scatter comes from genuine outliers to the relation \citep[see fig. 1 of ][]{2016ApJ...831..134V}; special systems which probably did not follow the same evolution as the majority of galaxies \citep[e.g.][]{2015ApJ...808...79F}. The same reservoir of gas that is fuelling AGNs or quasars, maintains the star-formation responsible for the growth of galaxies. This connection is evident from similar trends in the rates of star formation and black hole accretions with redshift \citep[e.g.][]{2004MNRAS.354L..37M,2014ARA&A..52..415M}, even though the actual growth of the black hole and the build up of the galaxy mass does not have to be concurrent, as the AGN duty cycles are relatively short and responses to the feedback are of different duration. 

Accretion models predict that black holes can reach masses of $10^{10}$ M$_\odot$ and such black holes have been detected at high redshift quasars \citep[e.g][]{2001AJ....122.2833F,2003AJ....125.1649F,2011Natur.474..616M,2015Natur.518..512W}. While such objects could explain the population of the highest mass black holes without the need for further growth via merging, they are not very common. Furthermore, there seems to exist a certain upper limit to the black hole growth via accretion of material \citep{2009MNRAS.393..838N}. The limit could be initiated by the feedback induced via mechanically or radiatively driven winds form the accretion disc around the black hole \citep{1998MNRAS.300..817H, 1998A&A...331L...1S, 1999MNRAS.308L..39F,2005ApJ...618..569M}. This means that the black holes with masses in excess to these predictions, if they exist, had to continue growing through mergers. 

Once when the AGN activity expels all gas, or the galaxy is massive enough and resides in a massive dark halo, which hinders the cooling of gas (and its accretion to the central black hole), the star formation is shut down and the mass growth of galaxies is possible only through dry merging. The accretion of small mass satellites changes the sizes of galaxies \citep[e.g.][]{2009ApJ...699L.178N}, but a significant increase of galaxy mass is only possible through major (similar mass) mergers. Such mergers are also characterised by eventual collisions of massive black holes (residing in the progenitors) and a fractional growth of black hole mass is equal to the fractional growth of galaxy mass. Still, the contribution of the mergers to the total growth of black holes has to be relatively small. \citet{2007ApJ...667..813Y} showed that mergers can change masses of black holes up to a factor of two, but this only happens in massive galaxies residing in galaxy clusters. \citet{2015ApJ...799..178K} confirmed this result showing that the black hole growth through mergers is only relevant for the most massive galaxies (and black holes), while the accretion is the main channel of growth. 

The channels of galaxy growth are essentially the same to those for black holes: one is dominated by the consumption of accreted gas in star-bursts and the other is dominated by accretion of mass through dissipation-less mergers. Numerical simulations show that these phases of growth can be separated in time \citep{2010ApJ...725.2312O}, where the early phase is characterised by in-situ formation of stars fuelled by cold flows \citep{2005MNRAS.363....2K,2009ApJ...703..785D}, while the later phase is dominated by accretion of stars formed elsewhere \citep{2012MNRAS.425..641L}. There is also a critical dependence on the mass of the galaxy as the less massive galaxies grow mostly with in-situ star formation, while only massive galaxies significantly grow by late stellar assembly \citep{2016MNRAS.458.2371R, 2017MNRAS.464.1659Q}. This supports the postulations that these are actually separate channels and galaxies follow either one or the other \citep{2016ARA&A..54..597C}.

The emerging paradigm of the growth of galaxies can be illustrated in the mass - size diagram \citep[see fig. 29 of ][]{2016ARA&A..54..597C}. For low mass galaxies, the redshift evolution of the distribution of galaxies in the mass - size diagram \citep{2014ApJ...788...28V} is explained by an inside-out growth of small star-forming galaxies. This phase of evolution increases the stellar mass within a fixed physical radius through star-formation, until the onset of quenching processes, which happen when galaxies reach a stellar density or a velocity dispersion threshold \citep{2015ApJ...813...23V}. At that moment galaxies transform from star forming spirals to fast rotator ETGs \citep{2013MNRAS.432.1862C}. This is characterised by a structural compaction of galaxies: a decrease in the size, and an increase in the concentration parameter (or the increase in the Sersic index of the light profiles), buildup of the bulge, and an increase in the velocity dispersion. While the quenching of star formation and buildup of bulges (or compacting processes) could be diverse \citep[e.g.][]{2014MNRAS.444.3408Y}, the main consequence is that galaxy properties related to the star formation history vary, on average, along lines of nearly constant velocity dispersion \citep[e.g.][]{2013MNRAS.432.1862C, 2015MNRAS.448.3484M,2016ARA&A..54..597C}. Given that the black hole growth is linked to the gas supply and the growth of the host, it is natural to expect that the black hole mass will closely follow the characteristic velocity dispersion of the host \citep{1998A&A...331L...1S,1999MNRAS.308L..39F}, where the details of the shape and scatter of the scaling relation depend on the details of the feeding of black holes \citep[e.g. how is the gas transported to the black hole,][]{2015ApJ...800..127A,2017MNRAS.464.2840A} and the feedback type \citep[e.g.][]{2005ApJ...618..569M}. The evidence shown in Fig.~\ref{f:ms2} supports a close relation between the star formation and the growth of black holes, as well as the cessation of star-formation and the final M$_{\rm BH}$. 

The picture is somewhat different for the very massive galaxies ($M_\ast \ga 2\times10^{11}$ M$_\odot$). They occupy a special region of the mass - size diagram; they are found along a relatively narrow ``arm'' extending both in mass and size from the distribution of other ETG \citep{2013MNRAS.432.1862C}. The narrowness of this arm is indicative of a small range in velocity dispersion that the most massive galaxies span, which is directly linked to the findings of \citet{2007ApJ...662..808L} and \citet{2007ApJ...660..267B} regarding the predictions for M$_{\rm BH}$ based on galaxy $\sigma$ or luminosity. This on the other hand provides a strong constraint on the processes that form massive galaxies: they have to increase both the mass and the size, but keep the velocity dispersion relatively unchanged. This particular property is characteristic for dissipation-less mergers of similar size galaxies \citep{1992ApJ...393..484B, 2003MNRAS.342..501N,2009ApJ...697.1290B, 2009ApJ...699L.178N}. Furthermore, as expected from such mergers, galaxies along this arm in the mass - size diagram have low angular momentum \citep{2011MNRAS.414..888E}, do not show evidence for containing stellar disks \citep{2011MNRAS.414.2923K}, harbour core-like profiles \citep{2013MNRAS.433.2812K}, are made of old stars \citep{2015MNRAS.448.3484M}, and are found at the density peaks of group or cluster environments \citep{2011MNRAS.416.1680C,2013ApJ...778L...2C, 2013MNRAS.436...19H, 2014MNRAS.441..274S,2014MNRAS.443..485F,2017arXiv170401169B}. For massive galaxies, the black hole growth can be linked to the growth of galaxies via dry mergers, and therefore, unlike for the low mass galaxies, it should be less strongly dependant on the velocity dispersion.

The expectation that growths of black holes and galaxies are connected implies that for star-formation driven build-up of galaxy mass (fuelled by direct accretion of gas or wet dissipational mergers) black holes grow by feeding, while in a merger driven growth of galaxies (via accretion of smaller objects to massive galaxies or major dry mergers) black holes will increase their mass through coalescence with other similar size black holes. The consequence is that at low galaxy masses the black holes should correlate with the galaxy velocity dispersion, while at high masses M$_{\rm BH}$ should be more closely related to the galaxy mass. The transition, however, is not sudden, and one can expect a persistence of the M$_{\rm BH} - \sigma$ relation to high galaxy masses. The reason for this is related to the expectation that high mass galaxies do not experience many similar mass dry mergers \citep[a few at most,][]{2009MNRAS.397..506K}. The significance of this is that only the most massive galaxies will go through a sufficient number of black hole mergers that increase M$_{\rm BH}$ disproportionally from the galaxy velocity dispersion. Therefore, the transition between the two regimes of M$_{\rm BH}$ dependence should be contingent on the galaxy mass, but it should be gradual and only visible at the highest masses, at a point beyond the critical mass of $2\times10^{11}$ M$_\odot$. 

There are challenges to this simple scenario already present in the literature. At low masses there are indications that different galaxy types follow different scaling relations with black hole mass when either bulge mass (luminosity) or effective velocity dispersion are taken into account \citep{2013ApJ...764..151G,2015ApJ...798...54G,2016ApJ...818...47S, 2016ApJS..222...10S}. At high masses, \citet{2015MNRAS.446.2330S} found that galaxies with most massive black holes reside in galaxies that seem to have undergone only a limited number of dissipation-less mergers (as measured by a proxy of the ratio between the missing light converted to mass and the black hole mass). It is, however, clear that low and high mass systems do go through different evolutionary paths, which might also be more diverse at lower masses. Confirming or disproving specific scenarios requires larger samples of galaxies with reliable black hole masses both at low and high galaxy mass range.

%
%

\section{Conclusions}
\label{s:con}

We used a recent compilation of black hole measurements, enhanced by a uniform determination of their sizes and total K-band luminosities, to show a variation of black hole masses in a mass - size diagram. As shown by previous studies \citep{2016ApJ...818...47S, 2016ApJ...831..134V}, black hole masses can be predicted from a combination of M$_\ast$ and R$_e$. In this study we show two additional characteristics of black hole masses. Firstly, black hole mass closely follows the changes in effective velocity dispersion in the mass - size plane, showing a similar behaviour as almost all properties of galaxies linked with star formation \citep{2016ARA&A..54..597C}. Secondly, there is tentative evidence that for higher masses (above $\approx2\times 10^{11}$ M$_\odot$) the black hole mass is progressively more correlated with the galaxy mass than with the velocity dispersion. 

We consider a physically motivated toy model in which black holes below a critical galaxy mass grow by accretion and follow M$_{\rm BH} - \sigma$ relation.  Above the critical mass black holes grow via mergers of similar size black holes. As these mergers are enabled by major dry mergers of galaxies, the black hole growth implicitly depends on the galaxy mass. As the critical galaxy mass, we choose M$_{\rm crit} = 2\times10^{11}$ M$_\odot$, which also roughly separates the regions in the mass - size plane populated by axisymmetric fast rotators and spiral galaxies from the slow rotators with cores in central surface brightness profiles \citep{2013MNRAS.432.1862C}. 

Assuming a M$_{\rm BH} - \sigma$ relation for Sersic galaxies \citep[][but other relations would also give similar results]{2013ApJ...764..151G}, our toy model has no free parameters and is able to qualitatively reproduce the trend in the data. While it does not provide a relation with less scatter than the standard M$_{\rm BH} - \sigma$ relation, it is physical motivated by the current paradigm of galaxy formation. The most massive galaxies, such as the central galaxies in groups and clusters, evolve through a different process than the bulk of the galaxy population. Namely, they experience multiple dissipation-less mergers, of which some (a few) are major and responsible for an equal increase of galaxy and black hole masses (through black hole binary mergers), but the stellar velocity dispersion remain unchanged. The consequence is a departure of black hole masses from the M$_{\rm BH} - \sigma$ relation for massive galaxies, in particular brightest cluster galaxies and massive slow rotators (with cores). This suggest that there should be a break in M$_{\rm BH} - \sigma$ and M$_{\rm BH} - $M$_\ast$ relations, similar to the one reported by \citet{2013ApJ...764..151G,2015ApJ...798...54G}, although the detection of the break or the need for more than a single power-law also depends on the choice of considered galaxies \citep[e.g. including or excluding galaxies of certain bulge type][]{2016ApJ...818...47S}. Irrespective of the chosen sample, the expected effect of the modulation of the M$_{\rm BH}$ is small as galaxies with suitable mass assembly history are rare and dry major mergers occur infrequently. 

The results presented here imply that there is no universal black hole - host galaxy scaling relation, but that it depends on the channel of formation that galaxies follow. The proposed model is simple and can easily be tested, but the black hole sample will have to include a larger number of massive and large galaxies than are currently available. The coming facilities such as JWST and E-ELT will allow us to reach such objects.

\section*{Acknowledgements}

DK acknowledges support from BMBF grant no. 05A14BA1 and thanks Alister Graham for pointing some relevant references in the literature. MC acknowledges support from a Royal Society University Research Fellowship. RMcD is the recipient of an Australian Research Council Future Fellowship (project number FT150100333). DK thanks Jakob Walcher for comments on an earlier version of the manuscript. 






\bsp	
\label{lastpage}
\end{document}